\documentstyle[preprint,pre,aps]{revtex}
\begin{document}
\draft
\small
\setcounter{page}{0}
\input epsf
\title{Categorization in fully connected multi-state neural network models}
\author{R. Erichsen Jr. and W. K. Theumann}
\address{Instituto de F{\'\i}sica,  Universidade Federal do Rio Grande do Sul,\\
  Caixa Postal 15051, CEP 91501-970, Porto Alegre, RS, Brazil}
\author{D. R. C. Dominguez}
\address{ESCET, Universidad Rey Juan Carlos, E 28933 Mostoles, Madrid, 
Spain and\\
CAB (Associate NASA Astrobiology)  INTA, 28850 Torrejon, Madrid, Spain}
\maketitle

\thispagestyle{empty}
\begin{abstract}
The categorization ability of fully connected neural network models, with
either discrete or continuous $Q$-state units, is studied in this work in
replica symmetric mean-field theory. Hierarchically correlated multi-state
patterns in a two level structure of ancestors and descendents (examples) are
embedded in the network and the categorization task consists in recognizing
the ancestors when the network is trained exclusively with their descendents.
Explicit results for the dependence of the equilibrium properties of a
$Q=3$-state model and a $Q=\infty$-state model are obtained in the form of
phase diagrams and categorization curves. A strong improvement of the
categorization ability is found when the network is trained with examples of
low activity. The categorization ability is found to be robust to finite
threshold and synaptic noise. The Almeida-Thouless lines that limit the
validity of the replica-symmetric results, are also obtained.
\end{abstract}


\label{sec:level1}

\section{Introduction}

Multi-state attractor neural networks in which the units (neurons)
can be in more than two states are, in general, more flexible and efficient
biological or artificial devices than networks of binary units. Much work has
been done over some time on the retrieval problem in multi-state networks of
various architectures, with either simple or hierarchical patterns in more
than two states. The retrieval problem consists in the recognition of patterns
that have been stored in a network by means of a learning (or training) rule,
when the network is set in an appropriate initial state to start its operating
stage \cite{HKP91}. Thus, the retrieval problem deals with
the memorization ability of a network. The networks that have been considered
are the dilute, the layered feed-forward and the fully connected networks
\cite{Ye89,MH89,R90,MKB,Ko,BE93,BRS94,KB,BH95,HBH,BE96,ET99,BE99,MW90,BV93}.

More recently, some work has been done on the categorization problem in
multi-state attractor networks \cite{ST92,DT95,DT96,DD96,DB97}, following
extensive studies of the problem in binary networks
\cite{FM89,F90,Mi91,BA,KT93,CT95,RF98,MT,LT,KT99}. 
The categorization problem
consists in the spontaneous recognition of a level of hierarchical patterns
other than those stored in the training process of a network \cite{FM89,F90}.
The problem deals with the ability to create a representation for concepts
when the network is only exposed to examples in the training stage.

Some of the questions that one may ask are the following. First, one is
interested in the minimal structure of the training patterns, and their
number, in order to achieve a satisfactory recognition of a macroscopic number
of hierarchically related ancestors. Second, one would like to know the
recognition rate (number of patterns per neuron) of these ancestors and how
stable they are as attractors of the network dynamics. The recognition quality
is of primary interest and one may also want to check on the robustness of the
recognition process to various kinds of noise.

The simplest, and most studied case of the categorization problem, consists in
the recognition of ancestors of a two-level hierarchy of ancestors and
descendents trained only with the latter according to a specific learning
rule. The hierarchical patterns that are generated trough a stochastic
procedure \cite{PV86} lead to correlations between patterns in
different levels as well as correlations between patterns in the same level
\cite{Gut}. As a consequence, there is a complex structure of attractors in
a network with hierarchical patterns in which the attractors may neither
coincide
with the training patterns nor with the ancestors, and it is of interest to
study under what conditions the latter become stable attractors.

Patterns in more than two states, which represent a gradual coding, may have a
low activity which is biologically appealing. Moreover, ``small''
patterns, in which a number of bits have been turned off, are patterns of low
activity that can infer patterns of full size and thereby enhance the
performance of a multi-state network, as demonstrated explicitly in works on
both the retrieval problem \cite{Ye89,MH89} and the categorization
problem. The dynamics of the latter has been studied in an extremely dilute
asymmetric three-state network with a monotonic neuron firing function and a
generalized Hebbian learning rule \cite{DT95}. The extremely dilute network
requires a
vanishingly small connectivity between neurons in order to allow for an exact
solution of the network dynamics, and one may ask what the behavior would be
for a network with full connectivity.

The purpose of the present paper is to answer some of the questions raised
above investigating the equilibrium, statistical mechanics behavior for the
categorization problem in a fully connected multi-state network with
hierarchical patterns of low activity in a two-level hierarchy of ancestors
and descendents. Our aim is to obtain the phase diagrams that describe the
various regimes of performance of the network in terms of the relevant
parameters: the activity of the training patterns, the dynamical activity of
the firing units, the correlation between ancestors and descendents, the
number of descendents, the multi-state threshold and the synaptic noise level,
assuming a fixed activity of the ancestors. The quality of the performance of
the network is described by so-called categorization curves that express the
dependence of the categorization error on some of the parameters of the
model. 

Since it is known from the results on the retrieval problem in a $Q$-state
network that the relevant phase diagrams become increasingly complex as one
goes from the three to the four-state model \cite{BRS94}, we
consider a $Q=3$-state model and a $Q=\infty$-state (graded response)
model. We make use of a generalized Hebbian learning rule that has been used
before \cite{DT95,DT96}. The outline of the paper is the following. In
section II we present the general $Q$ Ising-state model for the categorization
problem. The mean-field theory
for that model is summarized in section III. In section IV we present and
discuss the results for $Q=3$, in the absence
or presence of synaptic noise and in section V we present the results for
the $Q=\infty$-state model. We conclude in
section VI with a summary of the results.

\section{The model}

Consider a network of $N$ nodes, $i=1,\dots,N$. At the time step $t$, the 
state of the node $i$ is described by the variable $S_i(t)$, that can be in
any one of the $Q$ Ising states
\begin{equation}
\sigma_k=-1+\frac{2(k-1)}{Q-1}
\label{1}
\end{equation}
in the interval [-1,1], for $k=1,\dots,Q$. The task to be performed by the 
network is the recognition of a macroscopic set of $p$ concepts $\{\xi_i^{\mu};
\mu=1,\dots,p; i=1,\dots,N\}$, with $p=\alpha N$, where $\alpha$ is
finite. During the learning stage,
only a set of $s$ ``small'' examples $\{\xi_i^{\mu\rho}; \mu=1,\dots,p;
\rho=1,\dots,s;\, i=1,\dots,N\}$ of each concept are presented to the
network. By ``small'' examples we mean that a macroscopic number of bits in
each example
are turned off. The concepts are assumed to be independent identically
distributed random variables with zero mean and variance $A$. The examples
$\xi_i^{\mu\rho}$ of the concept $\xi_i^{\mu}$ are generated through a
stochastic process based on an appropriate probability distribution
$P(\lambda_i^{\mu\rho})$, given below, such that 
\begin{equation}
\xi^{\mu\rho}_{ i}=\xi^{\mu}_{i}\lambda^{\mu\rho}_{ i}\;.
\label{2}
\end{equation}

The properties of the distribution $P(\lambda_i^{\mu\rho})$ will be chosen in
accordance with the states of the neurons, Eq. (\ref{1}). For finite $Q=3$,
say $\lambda_i^{\mu\rho}$ assumes the values $+1$, $0$ or
$-1$ depending, respectively, on the example $\xi_i^{\mu\rho}$ being either in
agreement with the concept $\xi_i^{\mu}$, being turned off, or opposite
to the concept at the site $i$. In the case
of continuous neurons, i. e., $Q\rightarrow\infty$, we assume
that $\lambda_i^{\mu\rho}$ is a continuous variable in the interval
$[-1{,}1]$. In either case, we assume that $\lambda_i^{\mu\rho}$ belongs to a
set of independent random microscopic activities with mean
\begin{equation}
\langle\lambda_i^{\mu\rho}\rangle=b
\label{3a}
\end{equation}
and variance
\begin{equation}
\langle\lambda_i^{\mu\rho}\lambda_j^{\nu\sigma}\rangle
	=\left[b^2+\left(a-b^2\right)\delta_{\rho\sigma}\right]
\delta_{ij}\delta_{\mu\nu}
\label{3b}
\end{equation}
with $b^2\leq a\leq 1$. The symbol $\delta$ represents the Kronecker delta. In
consequence, we have the following relations,
\begin{equation}
\langle\xi_i^{\mu\rho}\xi_j^{\nu}\rangle
	=\langle\lambda_i^{\mu\rho}\xi_i^{\mu}\xi_j^{\nu}\rangle
	=bA\delta_{ij}\delta_{\mu\nu}
\label{4}
\end{equation}
and
\begin{equation}
\langle\xi_i^{\mu\rho}\xi_j^{\nu\sigma}\rangle
  =\langle\lambda_i^{\mu\rho}\lambda_j^{\nu\sigma}\xi_i^{\mu}\xi_j^{\nu}\rangle
	=[b^2+(a-b^2)\delta_{\rho\sigma}]A\delta_{ij}\delta_{\mu\nu}\;.
\label{5}
\end{equation}

The mean activity of the examples becomes 
\begin{equation}
\frac{1}{N}\sum_i^N(\xi_i^{\mu\rho})^2=aA\;,
\label{6}
\end{equation}
for every $\mu$ and $\rho$.
According to Eqs. (\ref{2}) and (\ref{3a}), $b$ is the correlation between 
an example and the concept to which it belongs. The pure multi-state model 
\cite{BRS94} can
be obtained by taking the number of examples $s=1$, the activity $a=1$ and 
the correlation $b=1$.
Since $a\leq 1$, the activity of the examples is not greater than the activity
of the concepts. In this sense, we refer to ``small'' examples, with the 
effective ``size'' of the patterns being $N_e=aN$. In this model, the view
point is that the small examples are samples of the full-activity concepts to
be inferred.

In this work we are interested in the capacity of the network to infer only
large concepts of full activity from the set of examples and restrict
ourselves, therefore, to binary concepts, $\xi_i^{\mu}=\pm 1$ with equal
probability, that is to say, to the case $A=1$. This
task is considered to be successful if the {\it categorization} overlap
\begin{equation}
m_{\mu}=\frac{1}{N}\sum_{i=1}^N\xi_i^{\mu}S_i
\label{6a}
\end{equation}
between the concept $\{\xi_i^{\mu}\}$ and the network state $\{S_i\}$
approaches unity after the network has reached the equilibrium state. 
To quantify the performance of the network, we define the categorization error
for the concept $\mu$ as
\begin{equation}
\varepsilon_c^{\mu}=\frac{1}{2}(1-m_{\mu})\,,
\label{6b}
\end{equation}
Thus, $\varepsilon_c^{\mu}$ should be small in the categorization phase and
$0.5$ in the disordered phase.

Next we pass to the discussion the dynamics of the model, following 
the steps of ref.~\cite{BRS94} and references therein. For a given 
configuration $\{S_i\}$ of the network, the local field $h_i$ on site $i$ is 
\begin{equation}
h_i(\{S_i\})=\sum_{j\ne i}J_{ij}S_j\;,
\label{7}
\end{equation}
where the synapses $J_{ij}$ are constructed from the examples, according to
the modified Hebb rule
\begin{equation}
J_{ij}=\frac{1}{N}\sum_{\mu=1}^p\sum_{\sigma=1}^s\xi_i^{\mu\sigma}\xi_j^{\mu
\sigma}\;{\rm for}\;i\ne j,\;J_{ii}=0.
\label{8}
\end{equation}
The state of each site is updated asynchronously according to a Glauber (single
spin-flip) dynamics in which the transition probabilities are given by 
\begin{equation}
P(S_j(t+\Delta t)=\sigma_k|\{S_i(t)\})
	\quad=\frac{\exp[\beta\epsilon_j(\sigma_k|h_j(\{S_i(t)\}))]}
	{\sum_{l=1}^Q\exp[\beta\epsilon_j(\sigma_l|h_j(\{S_i(t)\}))]}\;,
\label{9}
\end{equation}
where $\beta=1/T$ is the inverse temperature and the single site energy,
$\epsilon_j(s|h)$, is given by
\begin{equation}
\epsilon_j(s|h)=-hs+\theta s^2\,.
\label{10a}
\end{equation}
Here, $\theta$ is a non-negative constant that favors local states of small 
dynamical activity. In the absence of stochastic noise, the deterministic
evolution of the system is ruled by
\begin{equation}
S_j(t+\Delta t)={\rm dyn}(h_j(t)),
\label{10b}
\end{equation}
where ${\rm dyn}(x)$ is the non-decreasing step function, for finite $Q$,
\begin{equation}
{\rm dyn}(x)=\sum_{k=1}^Q\sigma_k[\Theta(\theta(\sigma_{k+1}+\sigma_k)-x)
		-\Theta(\theta(\sigma_k+\sigma_{k-1})-x)]
\label{11}
\end{equation}
with $\sigma_0=-\infty$ and $\sigma_{Q+1}=+\infty$, in which $\Theta(x)=1$, if
$x\geq 0$ and $0$ otherwise. The spin on site $j$ assumes the state $\sigma_k$
given by Eq.~(\ref{1}) if the local field $h_j$ is bound by
$\sigma_k+\sigma_{k-1}\leq h_j/\theta\leq\sigma_k+\sigma_{k+1}$. The width of
the intermediate states with constant $\sigma_k$ for $1<k<Q$ (that is,
excluding the limiting values of $\sigma_k=\pm 1$), is given by
$4\theta/(Q-1)$. Thus, the width of the zero state for the three-state network
studied below is $2\theta$. In the limit $Q\rightarrow\infty$, the
input-output function, Eq.~(\ref{11}), becomes the piecewise linear function
\begin{equation}
{\rm dyn}(x)={\rm sign}(x)
	{\rm min}\left(\left|\frac{x}{2\theta}\right|,1\right)\,,
\label{11a}
\end{equation}
where ${\rm min}(x,y)$ means the minimum between $x$ and $y$. The slope of the
linear part in here is $1/2\theta$, which is the gain parameter of the
continuous network.
The equilibrium thermodynamic properties of the fully connected infinite
network that follows from the above dynamics is described by the Hamiltonian
\begin{equation}
H=-\sum_{(ij)}J_{ij}S_iS_j+\theta\sum_{i}S_i^2\;,
\label{12}
\end{equation}
where the first sum is over all distinct pairs $(ij)$.

The relevant order parameters, when the network is in
the ordered sub-space of the phase space, are the {\it retrieval} overlaps
\begin{equation}
m_{\mu\rho}=\frac{1}{N}\sum_i^N\xi_i^{\mu\rho}S_i
\label{13}
\end{equation}
between the actual state of the network and each one of the examples $\rho$ of
each concept $\mu$. The underlying idea in studying the categorization
performance of the network is that when the number of correlated
examples is higher than a critical value, for a given correlation strength,
single examples are no longer local minima of the free-energy, but a mixed
state having macroscopic symmetric overlap $m_{\mu\rho}=m_s$, for
$\rho=1,\dots,s$,  with all the examples of a given concept $\mu$,
becomes a minimum. This state characterizes the categorization phase, and it
yields a finite, macroscopic, overlap $m_{\mu}$ with concept $\mu$.
Since we are interested mainly in the categorization ability of the network,
we restrict
ourselves, in what follows, to the study of configurations that have a
macroscopic overlap of order ${\cal O}(1)$ with a mixture of a finite number
$s$ of examples of
a given concept. Noting that the concepts are uncorrelated, one may
concentrate on the overlap with anyone of them, say $m_1$ for $\mu=1$.

\section{Mean-field theory}

The free-energy per site follows as
\begin{equation}
f(\beta)=-\lim_{N\rightarrow\infty}\frac{1}{\beta N}\left\langle
	\left\langle\ln {\cal Z}(\beta)\right\rangle_{\{\lambda^{\mu\rho}\}}
	\right\rangle_{\{\xi^{\mu}\}}\,,
\label{131}
\end{equation}
with the averages over examples and concepts in that order, as indicated,
where ${\cal Z}(\beta)$ is the canonical partition function
\begin{equation}
{\cal Z}(\beta)=\sum_{\{S_i\}}\exp(-\beta H)\,.
\label{132}
\end{equation}
In order to average over the quenched disorder, we employ the replica method,
in which 
\begin{equation}
\left\langle\left\langle\ln {\cal Z}(\beta)
	\right\rangle_{\{\lambda^{\mu\rho}\}}
	\right\rangle_{\{\xi^{\mu}\}}=\lim_{n\rightarrow 0}\frac{1}{n}
	\left(\left\langle\left\langle {\cal Z}^n(\beta)
	\right\rangle_{\{\lambda^{\mu\rho}\}}\right\rangle_{\{\xi^{\mu}\}}
	-1\right)\,.
\label{133}
\end{equation}
Using the generalized Hebb learning rule, Eq.~(\ref{8}), and introducing a
field $h_1$, in order to generate an equation for the overlap $m_1$, the
Hamiltonian Eq.~(\ref{12}), for the replica $a$, becomes
\begin{equation}
H^a=-\frac{1}{2N}\sum_{i\neq j}\sum_{\mu\rho}\xi_i^{\mu\rho}\xi_j^{\mu\rho}
	S_i^aS_j^a+\theta\sum_i(S_i^a)^2-h_1\sum_i\xi_i^1S_i^a\,.
\label{134}
\end{equation}
Introducing this expression in Eq.~(\ref{132}), separating the first
concept, we linearize the quadratic terms and obtain the replicated partition
function
\begin{eqnarray}
&&\!\!
\left\langle\left\langle\ln {\cal Z}(\beta)
	\right\rangle_{\{\lambda^{\mu\rho}\}}
	\right\rangle_{\{\xi^{\mu}\}}
	=\int\prod_{\alpha\rho}\frac{\sqrt{\beta N}{\rm d}m_{1\rho}^a}
	{\sqrt{2\pi}}\exp\left[-\frac{\beta N}{2}
	\sum_{a\rho}(m_{1\rho}^a)^2\right]\nonumber\\
	&&\quad\times
	\sum_{\{S_i^a\}}\left\langle\left\langle\exp(\beta pGn)
	\right\rangle_{\{\lambda_i^{\mu\rho}\}}\right\rangle_{\{\xi_i^{\mu}\}}
	\left\langle\left\langle\exp\left\{\beta\sum_{ia}\left[\sum_{\rho}
	m_{1\rho}^a\lambda_i^{1\rho}\xi_i^1S_i^a
	\right.\right.\right.\right.\nonumber\\
	&&\quad-\left.\left.\left.\left.
	\frac{1}{2N}\sum_{\rho}(\lambda_i^{1\rho}\xi_i^1
	S_i^a)^2-\theta(S_i^a)^2+h_1\xi_i^1S_i^a\right]\right\}
        \right\rangle_{\{\lambda_i^{1\rho}\}}\right\rangle_{\{\xi_i^1\}}\,,
\label{135}
\end{eqnarray}
where
\begin{equation}
\exp(n\beta pG)=\prod_{\mu>1}\left\langle\left\langle\exp\frac{\beta}
	{2N}\sum_{i\neq j}\sum_aS_i^aS_j^a
	\sum_{\rho}\lambda_i^{\mu\rho}\lambda_j^{\mu\rho}
	\xi_i^{\mu}\xi_j^{\mu}
	\right\rangle_{\{\lambda_i^{\mu\rho}\}}\right\rangle_{\{\xi_i^{\mu}\}}
	\,,
\label{136}
\end{equation}
involves the uncondensed examples.

In the thermodynamic limit,
$N\rightarrow\infty$ we obtain following ref.~\cite{IT},
\begin{equation}
n\beta G=-\frac{1}{2}{\rm tr}\ln\left(1-\beta\gamma_1\hat{Q}\right)
	-\frac{1}{2}(s-1){\rm tr}\ln\left(1-\beta\gamma_2\hat{Q}\right)
	-\frac{1}{2}\beta\gamma_1\,{\rm tr}\,\hat{Q}
	-\frac{1}{2}\beta(s-1)\gamma_2\,{\rm tr}\,\hat{Q}\,,
\label{139}
\end{equation}
where 
\begin{equation}
\gamma_1=a+(s-1)b^2\,,\;\;{\rm and}\;\;\gamma_2=a-b^2\,.
\label{15}
\end{equation}
Here, $\hat{Q}$ is a matrix in the space of replicas with elements given by
\begin{equation}
Q_{ab}\equiv\frac{1}{N}\sum_iS_i^aS_i^b=q_{ab}\;\;{\rm if}\;\;a\neq b\,,
\label{1310}
\end{equation}
and
\begin{equation}
Q_{aa}=Q_a\,.
\label{1311}
\end{equation}
Thus, $q_{ab}$ is the spin-glass order
parameter and $Q_a$ is the dynamical activity of the network. Whereas for the
binary network in the replica-symmetric theory $Q_a=1$, in the case of
multi-state networks one has, in general, that $q_{ab}\leq Q_a\leq 1$.

Introducing, as usual, the overlap parameter $r_{ab}$ associated to the
correlation between the overlaps of the examples and concepts that do not
condense, and restricting our study to the replica-symmetric solution, in which
\begin{eqnarray}
\begin{array}{c}
	m_{\mu\rho}^a=m_{\mu\rho}\\
	q_{ab}=q\\
	Q_a=a_D\\
	r_{ab}=r\,,\\
\end{array}
\label{1313}
\end{eqnarray}
we obtain that the replica-symmetric free-energy per site can be re-written as 
\begin{eqnarray}
f(\beta)&=&\frac{1}{2}\sum_{\rho=1}^s m_{1\rho}^2+\frac{\alpha}{2\beta}
	\left[\ln(1-\gamma_1C)+\frac{\gamma_1C}{1-\gamma_1C}
	+(s-1)\ln(1-\gamma_2C)+(s-1)\frac{\gamma_2C}{1-\gamma_2C}\right]
		\nonumber\\
	&+&\frac{\alpha}{2}\left[\frac{\gamma_1^2qC}{(1-\gamma_1C)^2}
	+(s-1)\frac{\gamma_2^2qC}{(1-\gamma_2C)^2}\right]
	-\frac{1}{\beta}\left\langle\left\langle\int{\cal D}z\,
	\ln\sum_{\{S\}}{\rm e}^{\beta{\cal H}_{eff}}\right\rangle
	_{\{\lambda^{1\rho}\}}\right\rangle_{\{\xi^1\}}\;,
\label{14}
\end{eqnarray}
where ${\cal D}z={\rm d}z\exp(-z^2/2)/\sqrt{2\pi}$ is a Gaussian measure.
The effective Hamiltonian, ${\cal H}_{eff}$, is given by
\begin{equation}
{\cal H}_{eff}=S\left(\sum_{\nu}m_{1\rho}\lambda^{1\nu}\xi^1
	-\theta'S+h_1\xi^1-\sqrt{\alpha r}z\right)\,,
\label{14a}
\end{equation}
where
\begin{equation}
\theta'=\theta-\frac{\alpha\gamma_1}{2(1-\gamma_1C)}
	-(s-1)\frac{\alpha\gamma_2}{2(1-\gamma_2C)}\,,
\label{16}
\end{equation}
is an effective width of the intermediate states, as wil be seen below (see
Eq.~(\ref{32})). Eventually, depending on the state of the network specified by
the dynamical activity 
$a_D$ and the spin-glass order parameter $q$, $\theta'$
may become negative, favoring an order with large absolute values for $S$.
Although Eqs.~(\ref{14})-(\ref{16}) follow from the
assumption of replica symmetry, we believe that such an order will exist, in
general, albeit in a small region of the phase space. Here,
$C\equiv\beta(a_D-q)=\beta\sum_i(\langle S_i^2\rangle-\langle S_i\rangle^2)/N$
represents the susceptibility of the network. The parameter $r$ is given by
the algebraic saddle-point equation
\begin{equation}
r=\frac{\gamma_1^2q}{(1-\gamma_1C)^2}
		+(s-1)\frac{\gamma_2^2q}{(1-\gamma_2C)^2}\,.
\label{17}
\end{equation}
The remaining saddle-point equations determining the order parameters are
\begin{equation}
m_{1\rho}=\left\langle\left\langle\lambda^{1\rho}\xi^1\int{\cal 
	D}z\,\langle S(z)\rangle\right\rangle_{\{\lambda^{1\rho}\}}
	\right\rangle_{\{\xi^1\}}\,,
\label{18}
\end{equation}
\begin{equation}
q=\left\langle\left\langle\int{\cal D}z\,
	\langle S(z)\rangle^2\right\rangle_{\{\lambda^{1\rho}\}}
	\right\rangle_{\{\xi^1\}}\;
\label{19}
\end{equation}
and
\begin{equation}
C=\frac{1}{\sqrt{\alpha r}}\left\langle\left\langle\int{\cal D}z\,
	z\langle S(z)\rangle\right\rangle_{\{\lambda^{1\rho}\}}
	\right\rangle_{\{\xi^1\}}\;.
\label{20}
\end{equation}
In the above equations,
\begin{equation}
\langle S^n(z)\rangle\equiv\frac{\sum_{\{S\}}S^n{\rm e}^{\beta{\cal H}_{eff}}}
	{\sum_{\{S\}}{\rm e}^{\beta{\cal H}_{eff}}}\,.
\label{23}
\end{equation}
The susceptibility $C$ remains finite, so that at zero temperature, that is
$\beta\rightarrow\infty$, $q\rightarrow a_D$, while at finite temperature we
have in general $q\leq a_D$. The overlap with the first concept, which
measures the categorization ability, is given by
\begin{equation}
m_1=\frac{\partial f}{\partial h_1}
	=\left\langle\left\langle\xi^1\int{\cal D}z\,
	\langle S(z)\rangle\right\rangle_{\{\lambda^{1\rho}\}}
	\right\rangle_{\{\xi^1\}}\;.
\label{21}
\end{equation}

Performing the configurational average in the saddle-point equations, we
obtain 
\begin{equation}
m_s=\frac{bm_1}{1-\gamma_2C}\,,
\label{25}
\end{equation}
for the symmetric overlap,
\begin{equation}
q=\int{\cal D}z\,S^2_{\beta}(h_s, \theta')
\label{26}
\end{equation}
and
\begin{equation}
C=\frac{1}{\sqrt{v}}\int{\cal D}z\,z\,S_{\beta}(h_s, \theta')\,,
\label{27}
\end{equation}
as well as the overlap with the concept,
\begin{equation}
m_1=\int{\cal D}z\,S_{\beta}(h_s, \theta')\,.
\label{28}
\end{equation}
The effective transfer function $S_{\beta}(h_s,\theta')$ is given by
\begin{equation}
S_{\beta}(h_s,\theta')=\frac{\sinh(\beta h_s)}{\frac{1}{2}{\rm e}^
	{\beta\theta'}+\cosh(\beta h_s)}
\label{29}
\end{equation}
in the case of the three-state network, and 
\begin{equation}
S_{\beta}(h_s,\theta')=\frac{h_s}{2\theta'}+\frac{1}{\sqrt{\beta\theta'\pi}}
	\frac{\exp\left[-\phi^2_+(h_s,\theta')\right]
	-\exp\left[-\phi^2_-(h_s,\theta')\right]}
	{{\rm erf}\left[-\phi_+(h_s,\theta')\right]
	-{\rm erf}\left[-\phi_-(h_s,\theta')\right]}\,,
\label{29a}
\end{equation}
where
\begin{equation}
\phi_{\pm}(h_s,\theta')=\sqrt{\beta\theta'}\left(1+\frac{h_s}{2\theta'}\right)
\label{29b}
\end{equation}
for $Q\rightarrow\infty$. Thus, $1/2\theta'$ is the effective gain parameter
for the continuous network. The effective field for the symmetric solution, 
$h_s$, is given by
\begin{equation}
h_s=sm_sb+z\sqrt{v}
\label{30}
\end{equation}
where
\begin{equation}
v=\alpha r+ sm_s^2\gamma_2\,.
\label{31}
\end{equation}
The first term in Eq.~(\ref{30}) is a signal term, while the second term is
the Gaussian noise due to the macroscopic number of uncondensed examples and
the presence of the symmetric mixture states. The latter, in which
$\gamma_2=a-b^2$ (cf. Eq.~(\ref{15})), is reduced in the case of examples of
low activity, $a<1$. One should expect, thus, an enhancement of the
categorization ability of the network in that case. The above equations are
obtained under the assumption that the number of examples $s$ is large, so
that the average over examples is given by a Gaussian distribution 
\cite{DT96,F90}. In the following sections we discuss the results based on
the solutions of the saddle-point
equations for both, the three-state and the continuous network.

The limit of stability of the replica-symmetric solution comes from the 
study of quadratic fluctuations of the free-energy in the vicinity of the
symmetric saddle-point. Following the Almeida and Thouless (AT) 
analysis~\cite{AT}, we obtain
\begin{equation}
\left(\frac{\gamma_1^2}{(1-\gamma_1C)^2}+(s-1)\frac{\gamma_2^2}
	{(1-\gamma_2C)^2}\right)
	\alpha\beta^2\left\langle\left\langle
	\int{\cal D}z\left[\langle S^2(z)\rangle-\langle S(z)\rangle^2\right]^2
	\right\rangle_{\{\lambda^{1\rho}\}}\right\rangle_{\{\xi^1\}}
	\leq 1\,.
\label{22}
\end{equation}
as the stability condition for the replica-symmetric solution.

\section{Three-state network}

\subsection{Categorization properties at zero temperature}

We begin by discussing the results for the categorization performance in
three-state networks in the absence of retrieval noise. The probability
distribution in this case is given by 
\begin{equation}
P(\lambda_i^{\mu\rho})=\frac{a+b}{2}\delta(\lambda_i^{\mu\rho}-1)
	+(1-a)\delta(\lambda_i^{\mu\rho})
		+\frac{a-b}{2}\delta(\lambda_i^{\mu\rho}+1)\,,
\label{3}
\end{equation}
satisfying the conditions (\ref{3a}) and (\ref{3b}). Thus, the
example $\xi_i^{\mu\rho}$ has a probability $(a+b)/2$ to be aligned with
the concept, while it has a probability $1-a$ to be turned off and a
probability $(a-b)/2$ to be opposed to the concept. 

The effective transfer function, Eq.~(\ref{29}), at zero temperature becomes
\begin{equation}
S_{\infty}(h_s,\theta')\equiv\lim_{\beta\rightarrow\infty}
S_{\beta}(h_s,\theta')={\rm sgn}(h_s)\Theta(|h_s|-\theta')\,.
\label{32}
\end{equation}
From Eq.~(\ref{16}), we see that $\theta'$ may become negative. Since
$S_{\infty}(h_s,\theta'<0)$ is algebraically the same as 
$S_{\infty}(h_s,\theta'=0)$ the network acts, in this case, as a binary
network at $T=0$. Accordingly,
Eq.~(\ref{25}) remains unchanged, while Eqs.~(\ref{26}) and (\ref{27})
become
\begin{equation}
q=1-\frac{1}{2}{\rm erf}\left(\frac{sm_sb+\theta'\Theta(\theta')}
	{\sqrt{2v}}\right)
	+\frac{1}{2}{\rm erf}\left(\frac{sm_sb-\theta'\Theta(\theta')}
	{\sqrt{2v}}\right)
\label{33}
\end{equation}
and
\begin{equation}
C=\frac{1}{\sqrt{2\pi v}}\exp\left[-\frac{(sm_sb+\theta'\Theta(\theta'))^2}
	{2v}\right]+\frac{1}{\sqrt{2\pi v}}
	\exp\left[-\frac{(sm_sb-\theta'\Theta(\theta'))^2}{2v}\right]\,.
\label{34}
\end{equation}
The overlap with the concept, Eq.~(\ref{28}), is given by
\begin{equation}
m_1=\frac{1}{2}{\rm erf}\left(\frac{sm_sb+\theta'\Theta(\theta')}
	{\sqrt{2v}}\right)
	+\frac{1}{2}{\rm erf}\left(\frac{sm_sb-\theta'\Theta(\theta')}
	{\sqrt{2v}}\right)\,.
\label{35}
\end{equation}

We show in Fig.~\ref{fig1} the categorization phase-diagram 
for the case where $s=20$, $a=0.2=b$. With the choice that $a=b$, we are
looking in a way for an optimal phase diagram in the sense that the training
examples either coincide with the corresponding concepts, that is
$\xi_i^{\mu\rho}=\xi_i^{\mu}$, for $\rho=1,\dots,s$, or are zero, but they are
never opposed to the concept. For other values of the
parameters, similar diagrams are obtained, although with lower capacity
$\alpha$. The categorization phase (C), characterized by
$m_1\neq 0$ and $q\neq 0$ is globally stable below the heavy solid line. It
becomes only locally stable, while
the spin-glass (SG) phase is globally stable, between the heavy
solid and the light solid line, where the system always jumps
discontinuously to the spin-glass phase. This is in distinction with known
results for the categorization phase diagram in the dilute
network~\cite{DT96}, where the transition to the spin glass phase is partly
continuous and partly discontinuous. Above the light solid line, and at the
left 
of the dash-dotted line, where it disappears continuously, the spin-glass 
phase, with $m_1=0$ and $q\neq 0$, is stable. At the right of the dash-dotted
line the paramagnetic (P), or {\sl zero}-phase, with $m_1=0$ and $q=0$, is
stable. Note that, for large threshold $\theta$, there is a direct transition
from the categorization phase to the fully disordered P phase, at low
$\alpha$. There exists also a retrieval phase of examples, without
categorization, not shown in the figure. Since we are dealing with a large
number of examples (thus favoring the categorization), that phase is present
in the phase diagram only at very small values of $\alpha$ and $\theta$.
To the left of the dotted line, the
effective width $\theta'$ is negative. Here, every non-zero value of the local
field is sufficient to access the neural states $S_i=\pm 1$ and, in
consequence, the network behaves in this region as a binary network. The dashed
line signals the optimal $\theta$, i.e., the value of the width parameter for
which the categorization overlap $m_1$ reaches its maximum value. 
It is interesting to note that the present phase diagram is similar to that of
ref.~\cite{BRS94}, for the retrieval problem, with the categorization phase
taking the role of the retrieval phase in that problem. 

An important question addressed in this paper refers to the role played by the 
activity of the examples, $a$, on the categorization ability of the
network. In Fig.~\ref{fig2} the categorization error $\varepsilon_c$ is shown as
a function of the activity, for $\alpha=0.02$, 
$s=20$, $b=0.2$ and for several values of $\theta$. The results
reveal that $\varepsilon_c$ is a monotonically increasing function of $a$. Since
this is the general behavior for other values of the parameters, the results
confirm that for the connected, as well as for the 
dilute~\cite{DT96} networks, it is better to train the network with
low-activity examples. This can be understood noting that the activity $a$
of the examples is decreased when a macroscopic number of bits of every
example is turned off. But in keeping the overlap $b$ between examples and
concepts fixed, the bits that are turned off in the examples must be those that
are inverted with respect to the concepts . When the activity $a$ reaches its
minimal, optimal value, $a=b$, the only bits that are turned on in
the examples are those that are aligned with the concepts, leading to the
smallest categorization error. In this case the categorization task of the
network becomes similar to the reconstruction of a puzzle from loose
pieces. Finally, the figure also shows the discontinuous jump to the
spin-glass (SG) phase, at the upper phase boundary of Fig.~\ref{fig1}.

The categorization error as a function of the number of 
examples $s$, for $b=0.4$, $\theta=0$ and $\theta=1.0$ and two different
activities, namely $a=b$ (all wrong bits in the examples are turned off) and
$a=1.0$ (all wrong bits in the examples are included) is shown in 
Fig.~\ref{fig3}. Starting from the spin glass phase, with 
categorization error equal to $0.5$, the network
undergoes a discontinuous transition to the categorization phase as the number 
of examples increases above a critical value. The number of examples required
for the jump to the categorization phase is considerable smaller for $a=0.4$,
than for $a=1.0$. Nevertheless, the final categorization error is similar for
the two activities. This means that the
network is able to overcome a higher amount of errors in the examples by a
larger number of these examples. A higher value of $s$ is also required
for a higher threshold $\theta$, in order to reach a higher local field to
attain the states with non-zero activity.

\subsection{Categorization properties in the presence of synaptic noise}

We consider next the categorization performance obtained from
$S_{\beta}(h_s,\theta')$, Eq.~(\ref{29}), for finite $\beta$. Fig.~\ref{fig4}
illustrates the influence of the temperature on the categorization error, for 
$\alpha=0.01$, $s=20$, $b=0.2=\theta$, and activity $a$ equal to $0.2$ and 
$0.3$. To the left of the arrow in the curve corresponding to $a=0.2$, the
categorization phase is the global minimum, while it is a local minimum to the
right. In what concerns the present set of parameters, the categorization
phase for $a=0.3$ is a local minimum for all temperatures, whereas the
spin-glass phase is the global minimum. Thus, it is also advantageous for
an enhancement of the performance of the network, in the presence of synaptic
noise, to train the network with examples of low activity. Fig.~\ref{fig4}
also shows the discontinuous transition to
the spin-glass phase at an activity-dependent transition temperature. 

The phase diagram for $\alpha$ vs. $T$ is presented in Fig.~\ref{fig5} for 
$\theta=0.2$, $s=20$ and $a=0.2=b$. The categorization phase is stable below
the upper phase boundary, where it disappears discontinuously, becoming a
global minimum below the lower phase boundary. At very small $\alpha$ and $T$
there is a retrieval phase without categorization, not shown in the
figure. The dashed line on the left is the 
locus of the AT-line. The replica-symmetric solution for the categorization 
phase becomes unstable to replica-symmetry-breaking fluctuations at the left of
this line. The re-entrant behavior of the upper phase boundary at low $T$ is
associated to the instability of the replica-symmetric solution in this
region. The spin-glass phase becomes a global minimum to the right of the
heavy solid, and to the left of the dotted line, where it disappears
continuously. At the right of the dotted line, the paramagnetic phase is the
global minimum.

\section{Network with continuous neurons}

In this section we discuss the categorization properties of a network with 
continuous, monotonic neurons trained with continuous or discrete examples of
binary concepts. The continuous limit is
obtained by taking $Q\rightarrow\infty$ in Eqs.~(\ref{1}) and (\ref{11}). 
The following results are independent of the specific form of 
$P(\lambda_i^{\mu\rho})$, provided that its mean and variance are given by
Eqs.~(\ref{3a}) and (\ref{3b}), respectively. The general
Eqs.~(\ref{25})-(\ref{28}), for the saddle-points, apply also to this case. In
the absence of noise, the effective transfer function, Eq.~(\ref{29a}),
becomes the stepwise linear function
\begin{equation}
S_{\infty}\left(h,\theta'\right)={\rm sign}\left(\frac{h}{2\theta'}\right)
	{\rm min}\left(\left|\frac{h}{2\theta'}\right|,1\right)\,,
\label{41}
\end{equation}
in which $1/2\theta'$ is the effective gain parameter.
Consequently we obtain
\begin{eqnarray}
m_1&=&\frac{1}{2}\left(1-\frac{sm_sb}{2\theta'}\right){\rm erf}
	\left(M_-\right)
	+\frac{1}{2}\left(1+\frac{sm_sb}{2\theta'}\right)
	{\rm erf}\left(M_+\right)\nonumber\\
& &\quad\quad-\frac{1}{2\theta'}\sqrt{\frac{v}{2\pi}}\left[\exp\left(-M_
	-^2\right) 
	-\exp\left(-M_+^2\right)\right]\,,
\label{42}
\end{eqnarray}
\begin{eqnarray}
q&=&1-\frac{v}{4\theta'^2\sqrt{\pi}}\left[M_+
	\exp\left(-M_-^2\right)-M_-
	\exp\left(-M_+^2\right)\right]
	\nonumber\\
& &+\frac{1}{2}\left[1+\frac{v}{2\theta'^2}\left(\frac{1}{2}
	+\frac{s^2m_s^2b^2}{2v}\right)\right]\left[{\rm erf}\left(M_-\right)
	-{\rm erf}\left(M_+\right)\right]
\label{43}
\end{eqnarray}
and
\begin{equation}
C=\frac{1}{4\theta'}\left[{\rm erf}\left(M_+\right)
	-{\rm erf}\left(M_-\right)\right]\,,
\label{44}
\end{equation}
where
\begin{equation}
M_{\pm}=\frac{sm_sb\pm 2\theta'}{\sqrt{2v}}\,.
\label{441}
\end{equation}

The zero-temperature phase diagram for $\alpha$ vs. $\theta$ with $s=20$ and
$a=0.2=b$ is shown in
Fig.~\ref{fig6}. The categorization phase exists below the light solid
line, and it is the global minimum below the heavy solid line. At the left of
the dotted line, $\theta'$ is zero and the effective gain is infinite. In
this region, the states $S_i=\pm 1$ are the only accessible states for
non-zero local field and the network behaves as a binary network. In there,
the critical $\alpha$ for categorization assumes its value in the binary
network for this set of parameters, i. e.,
$\alpha_{c,binary}\approx 0.033$. When the network enters the multi-state,
continuous regime, the categorization capacity starts to increase abruptly,
and reaches its maximum value $\alpha_c\approx 0.047$ for $\theta\approx 0.11$.
The dashed line signals the
optimum $\theta$ for each $\alpha$. It is worth noting that for
$\alpha<\alpha_{c,binary}$ the optimal $\theta$ line coincides with the
transition to the binary regime. This means that whenever there is a binary
network capable to perform the categorization task, it will give the best
categorization properties for low $\theta$. Only when
$\alpha>\alpha_{c,binary}$ the network with continuous neurons is expected to
have a better performance. Contrary to the case of finite $Q$,
where at zero temperature the replica-symmetric solution is always unstable,
there is here a region where it is stable, and this is the part of the phase
diagram below the light dash-dotted line. The phase diagram illustrates that
also the network of continuous neurons is robust to low gain in the
states. 
The
existence of a replica-symmetric stable phase at zero temperature was noticed
in ref.~\cite{BRS94}, for the retrieval problem in a network of continuous
neurons. Finally, the heavy dash-dotted line represents the onset of the
continuous spin glass transition.

In Fig.~\ref{fig7} we present the categorization error as a function of the
activity of examples, for $\alpha=0.02$, $s=20$, and $\theta=0.2$ to
$0.4$. Since we deal with a non specified $P(\lambda_i^{\mu\rho})$, the only
restriction imposed is $a\geq b^2$. We note from the figure that the
categorization 
error is no longer a monotonic increasing function of the activity for all
values of $\theta$. For $\theta=0.4$, $\varepsilon_c$ is 
a decreasing function of $a$, for small $a$. The reason is that in the case of
large threshold $\theta$, the local field $h_i(\{S_i\})$ must be
sufficiently high to overcome the threshold, and this is obtained through a
moderate increase in the activity of the examples.

Finally, we discuss the influence of the number of examples $s$ on the
categorization ability of networks with continuous neurons. Fig.~\ref{fig8}
shows the
categorization error as a function of $s$ for $a=0.2=b$, threshold ranging
from $0.2$ to $0.4$ and $\alpha=0.02$. As a result of the continuous nature of
the units, for low threshold the categorization error decreases smoothly with
the increasing number of examples. This is distinct to the previous case of
discrete units, where an abrupt decrease in $\varepsilon_c$ was observed even at
$\theta=0$ (see Fig.~\ref{fig1}). Furthermore, the
decreasing in $\varepsilon_c$ is no longer monotonic for all values of the
threshold. For example, for $\theta=0.2$ there is a local maximum in
$\varepsilon_c$ for $s\approx 30$.

\section{Summary and concluding remarks}

The categorization problem, that consists of the recognition of ancestors,
when a network is trained only with their descendents, is studied in this work
for multi-state fully connected neural network models, keeping in mind an
application to either artificial or biological networks in which the training
is with sparsely coded patterns. Indeed, multi-state
networks offer the possibility of recognizing full-sized patterns in networks
trained with ``small'' patterns, in which a macroscopic number of bits have
been reduced or, eventually, set to zero reducing thereby the activity of the
encoded patterns. We found that a low activity can enhance the categorization
ability of a fully connected network in a significant way, by changing the
threshold for firing of the units. This  confirms and extends earlier results
on an extremely dilute network of $Q=3$-state neurons \cite{DT95}.

The way the network works for the categorization task is the following. After
training with correlated examples, the network searches for stable symmetric
mixtures states, in place of pure examples. If these patterns have low
activity, it will be less likely that they have bits with opposite sign to
the corresponding concepts. The recognition of the latter from the common
features of the examples will thereby be enhanced.

We derived formal expressions, within replica-symmetric mean-field theory, for
the free energy and the relevant order parameters for the categorization
problem in a fully connected neural network model, with units in general $Q$
Ising-states and multi-state patterns belonging to a two-level
hierarchy. Training of the network was assumed to take place through a
generalized Hebbian learning rule involving only the descendents. These may be
considered as corrupted examples of the ancestors (concepts) with a number of
turned off or inverted bits. Explicit results for the relevant phase diagrams
and the categorization curves were then obtained for a $Q=3$-state model with
a monotonic activation function and for a monotonic $Q=\infty$-state model. In
the first case we also checked the robustness of the network performance to
synaptic noise. Our results are restricted to binary ancestors and multi-state
descendents, although the case of multi-state ancestors has been considered in
an extremely dilute network \cite{DB97}.

The limit of validity of the replica-symmetric solution was established
in this work looking for the Almeida-Thouless lines. For $Q=3$, the
replica-symmetric solution is unstable in the absence of synaptic noise ($T=0$)
and there is a re-entrant behavior for the ratio $\alpha$ of recognized
concepts, at small synaptic noise, in accordance with earlier results on the
retrieval problem \cite{CN92,NC92} and on the
categorization problem in connected networks of binary neurons
\cite{KT99}. Nevertheless, since the replica-symmetric solution stabilizes at
very small $T$, we argue that replica-symmetry breaking effects should be
negligible,
even at $T=0$. On the other hand, there is a finite region of interest for the
categorization performance domain where the replica-symmetric solution is
stable, even at $T=0$, in the case of the $Q=\infty$-state network, as
demonstrated explicitly in this work.

To summarize, we succeeded in studying a fully connected multi-state neural
network model for the categorization problem of recognizing binary concepts
when the network is trained with $Q$-state examples of low activity, in place
of the full activity patterns of a binary network of states $S=\pm 1$. The
work presented here can be extended in various directions. First, to infer
multi-state concepts in a network with full connectivity and to study the
categorization performance for sparsely coded sequential examples. In order to
come closer to biological networks, it would be interesting to consider the
partial dilution of synapses.

{\bf Acknowledgments}

We are indebted to Alba Theumann for showing us how to find the
Almeida-Thouless line for a network with hierarchically correlated
patterns. We thank D. Boll{\'e} for comments and discussions, as well as for
the kind hospitality of the Institute for Theoretical Physics of the Catholic
University of Leuven, where part of the work of DRCD and WKT was done.
This work was supported, in part, by CNPq 
(Conselho Nacional de Desenvolvimento Cient{\'\i}fico e Tecnol{\'o}gico,
Brazil), and FINEP (Financiadora de Estudos e Projetos, Brazil).

\newpage

\begin{figure}
\caption{Phase diagram for the ratio $\alpha$ of recognized concepts as a
function of the threshold $\theta$, for the three-state network. The number of
examples is $s=20$, the activity $a=0.2=b$ (the correlation parameter). Here,
C, SG and P are the categorization, spin-glass and paramagnetic phases,
respectively. Below the heavy solid line, the categorization phase is the
absolute minimum of the free-energy. Solid (dash-dotted) lines indicate a
discontinuous (continuous) transition. The dashed line indicates the optimal
value of $\theta$. At the left of the dotted line, the network behaves as a
binary network with states $S_i=\pm 1$.}
\label{fig1}
\end{figure}

\begin{figure}
\caption{Categorization error as a  function of the activity $a$, for the
three-state network at $T=0$, when $\alpha=0.02$,
$s=20$, $b$=0.2 and $\theta=0.0$--$0.3$ (as indicated).}
\label{fig2}
\end{figure}

\begin{figure}
\caption{Categorization error as a function of the number of examples $s$, for
the three-state network at $T=0$, when $\alpha=0.05$, $b=0.4$. Solid (dotted)
lines correspond to $\theta=0.0$
($\theta=1.0$) and the two lines at the left (right) correspond to $a=0.4$
($a=1.0$).}
\label{fig3}
\end{figure}

\begin{figure}
\caption{Categorization error as a function of the temperature $T$, for the
three-state network, when $\alpha=0.01$, $s=20$, $b=0.2=\theta$ and $a=0.2$
(solid line) and $0.3$ (dotted line).}
\label{fig4}
\end{figure}

\begin{figure}
\caption{Categorization phase diagram of $\alpha$ vs. $T$, for the
three-state network, when  $\theta=0.2$, $s=20$ and $a=0.2=b$. Below  
the heavy solid line the categorization phase is the absolute minimum of the
free energy. Solid (dotted) lines indicate a discontinuous (continuous)
transition. The replica symmetry is broken at the left of the dashed line.}
\label{fig5}
\end{figure}

\begin{figure}
\caption{Categorization phase diagram of $\alpha$ vs. $\theta$, for the
continuous $Q=\infty$-state network, at $T=0$, when $s=20$, $a=0.2=b$. Here,
C, SG and P are the categorization, spin-glass and paramagnetic phases,
respectively. Below the heavy solid line the categorization phase is the
absolute minimum of the free-energy. Solid (heavy dash-dotted) lines indicate
discontinuous (continuous) transitions. The dashed line indicates the optimal
value of $\theta$. At the left of the dotted line, the network behaves as a
binary network with states $S_i=\pm 1$. Below the light dash-dotted line the
replica symmetric solution is stable.}
\label{fig6}
\end{figure}

\begin{figure}
\caption{Categorization error as a function of the activity $a$, for the
$Q=\infty$-state network at $T=0$, when $\alpha=0.02$, $s=20$ and $b$=0.2. The
threshold values are $0.2$, $0.3$ and $0.4$ (curves from right to
left).} 
\label{fig7}
\end{figure}

\begin{figure}
\caption{Categorization error as a function of the number of examples $s$, for
the $Q=\infty$-state network at $T=0$, when $\alpha=0.02$ and $a=0.2=b$. The
threshold values are $0.2$, $0.3$ and $0.4$ (curves from left to right).}
\label{fig8}
\end{figure}

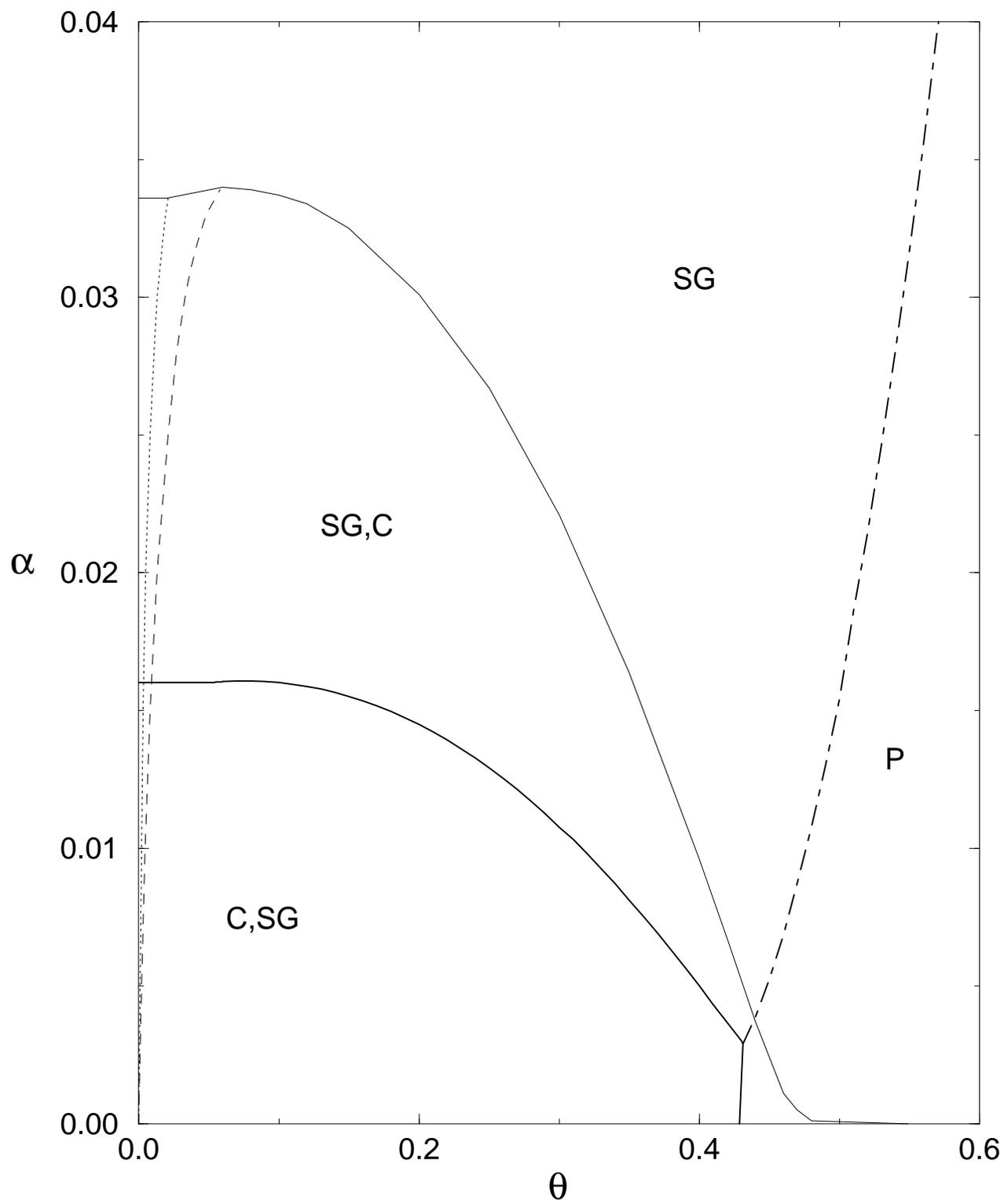
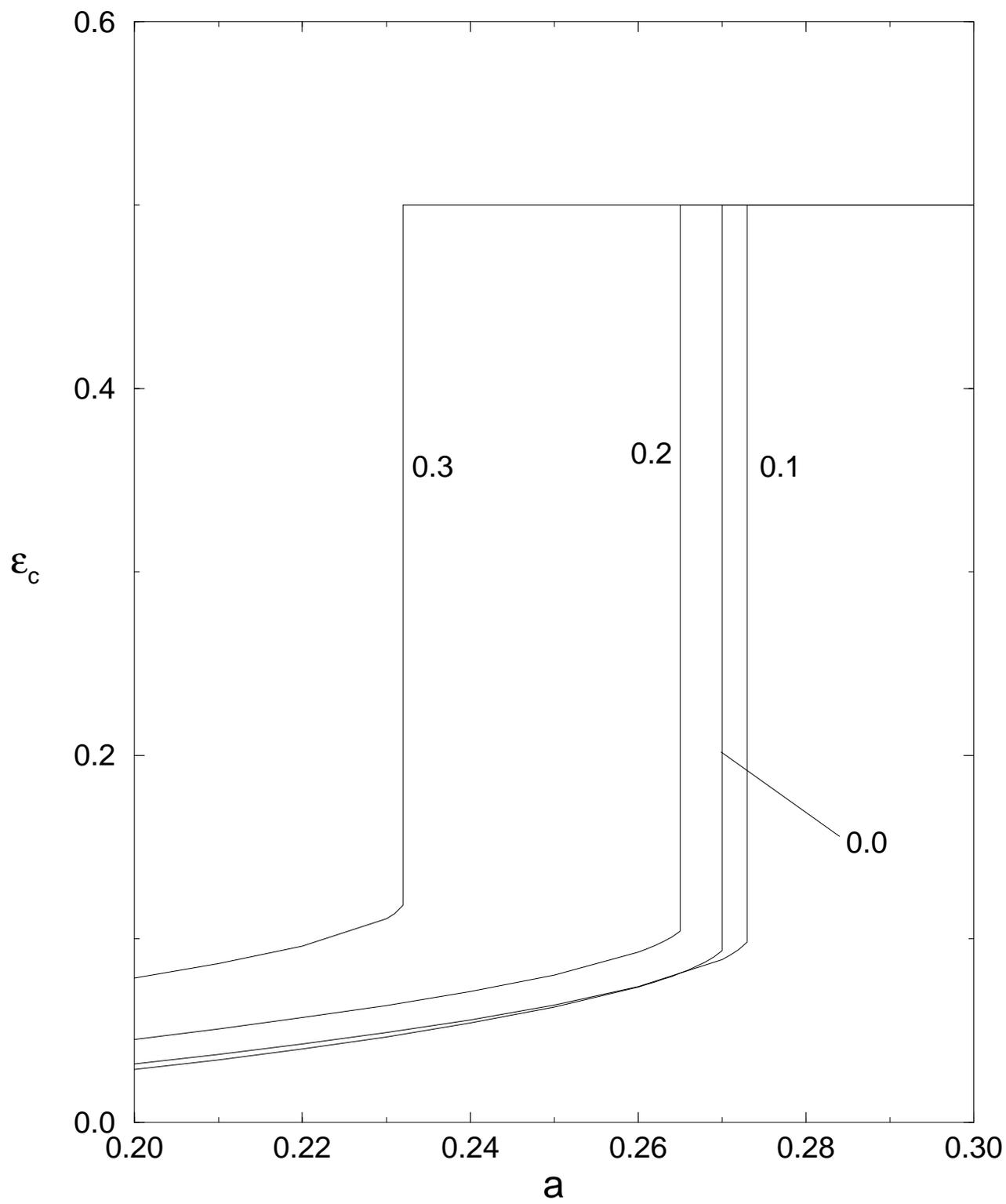
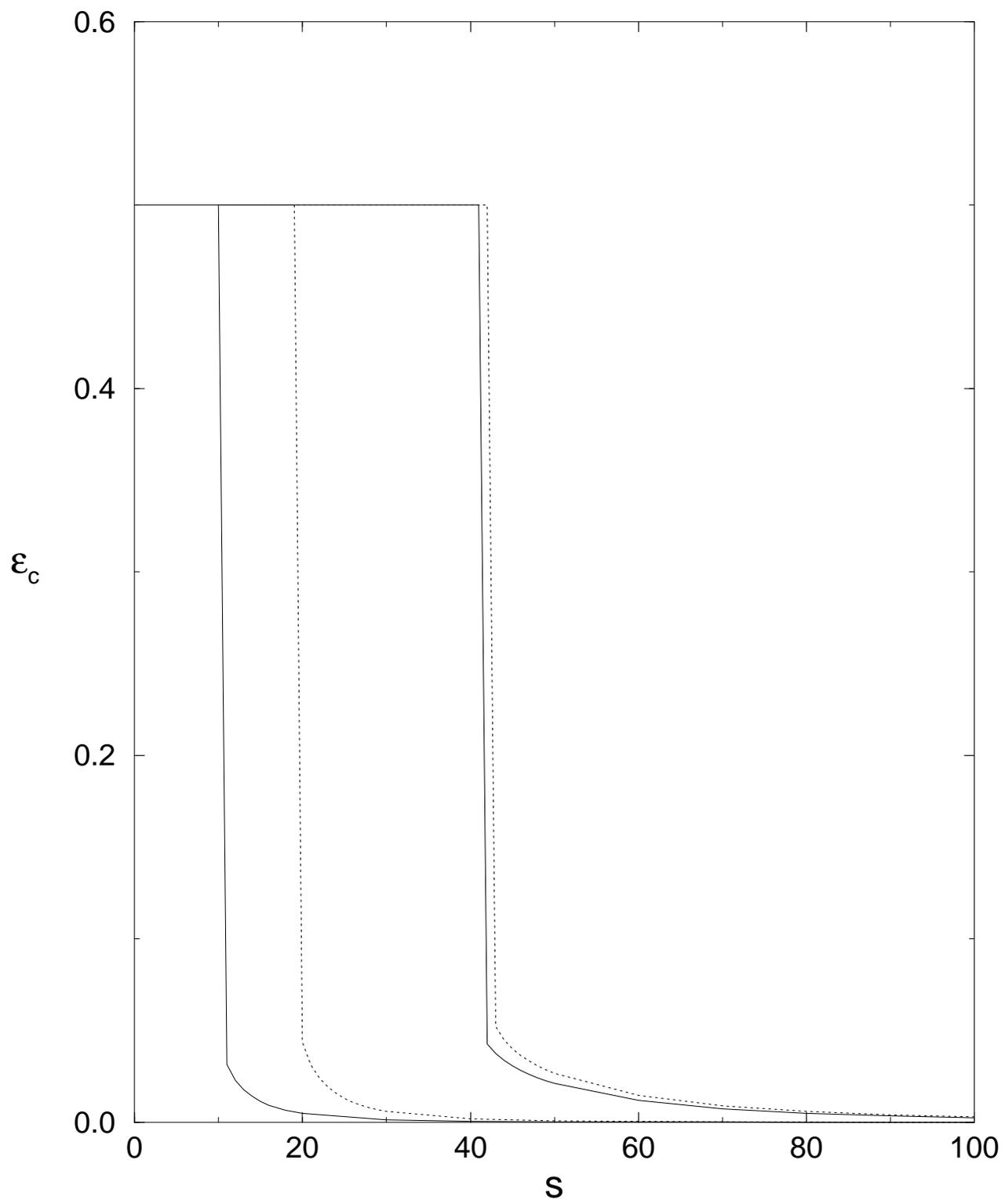
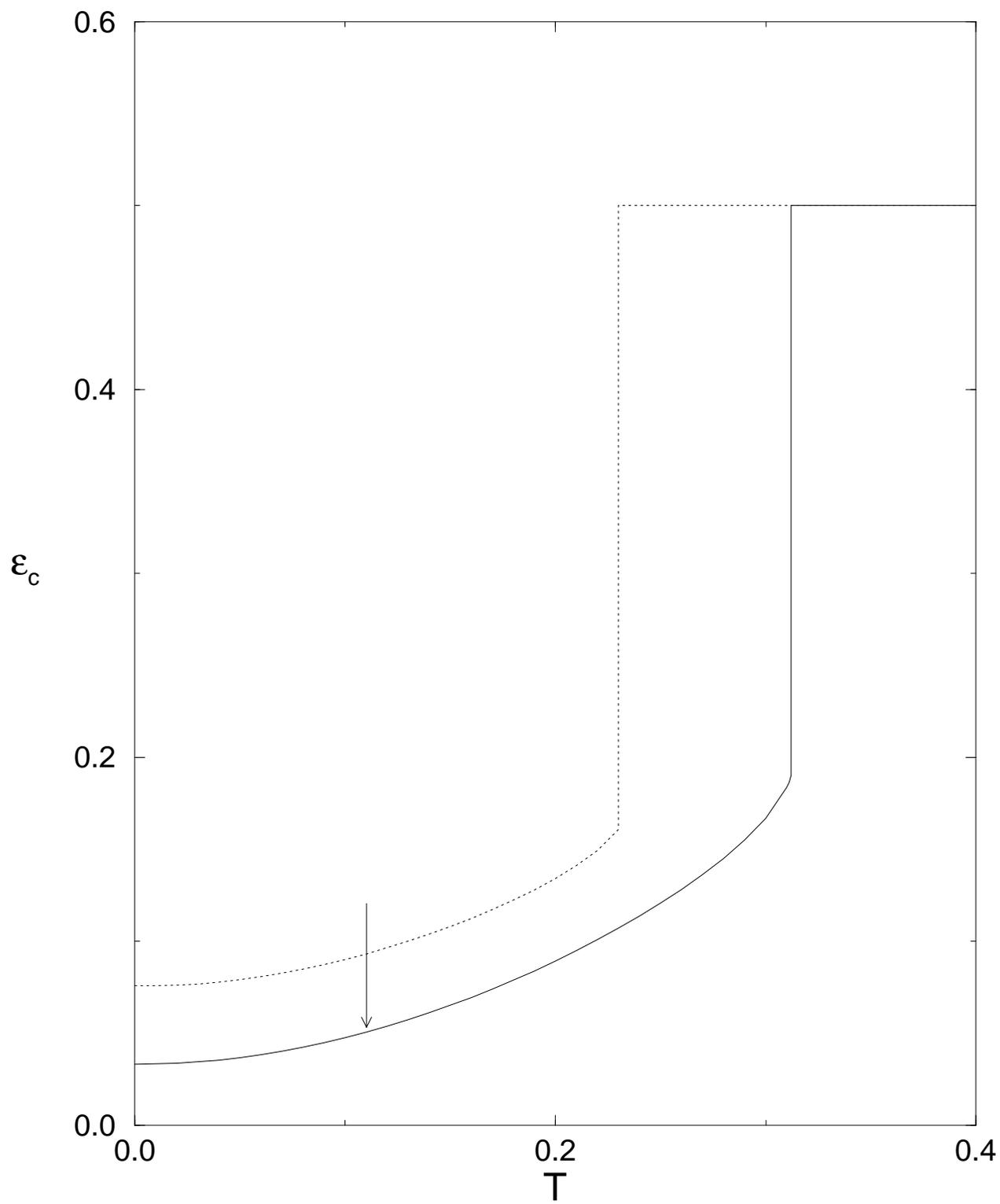
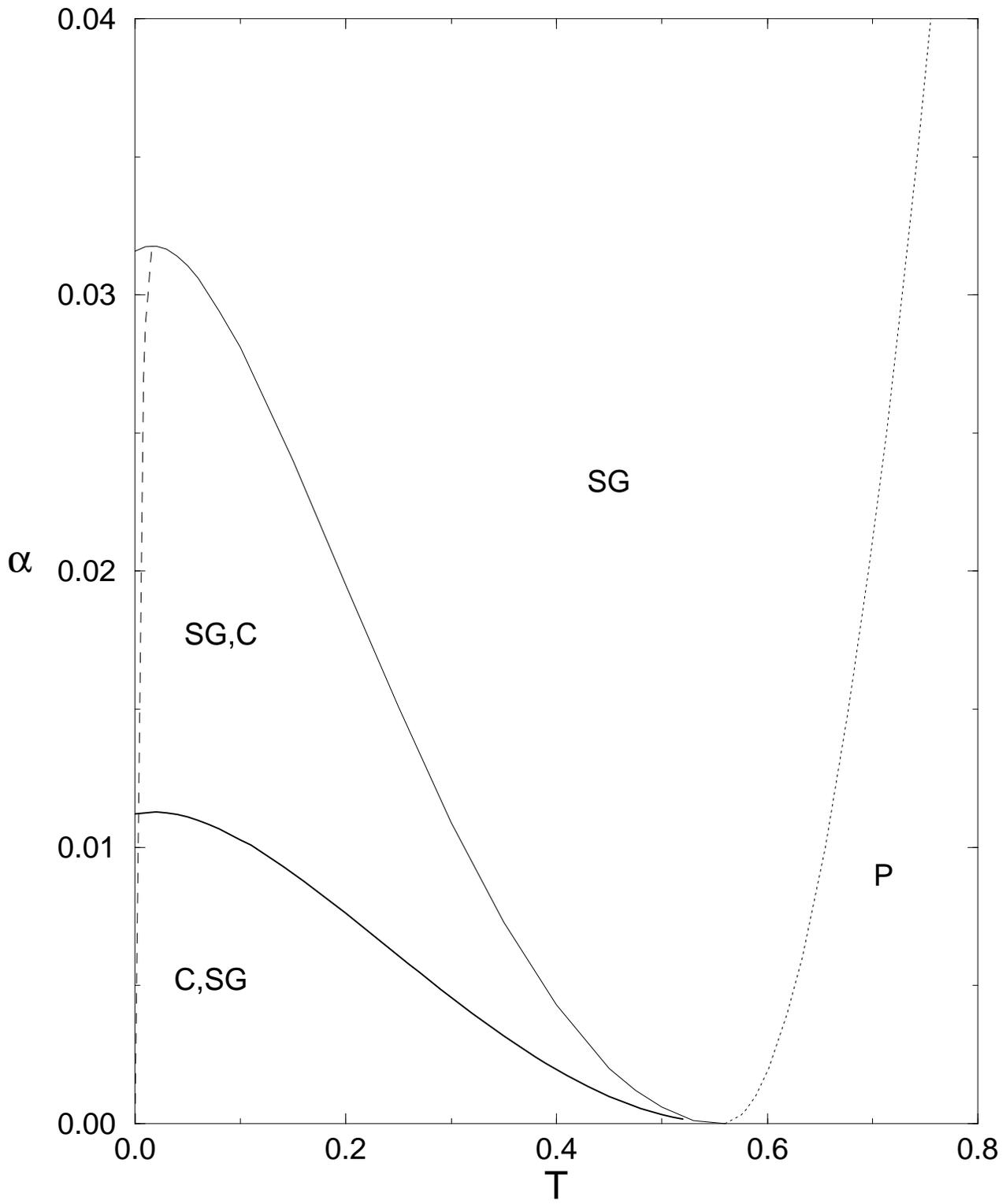
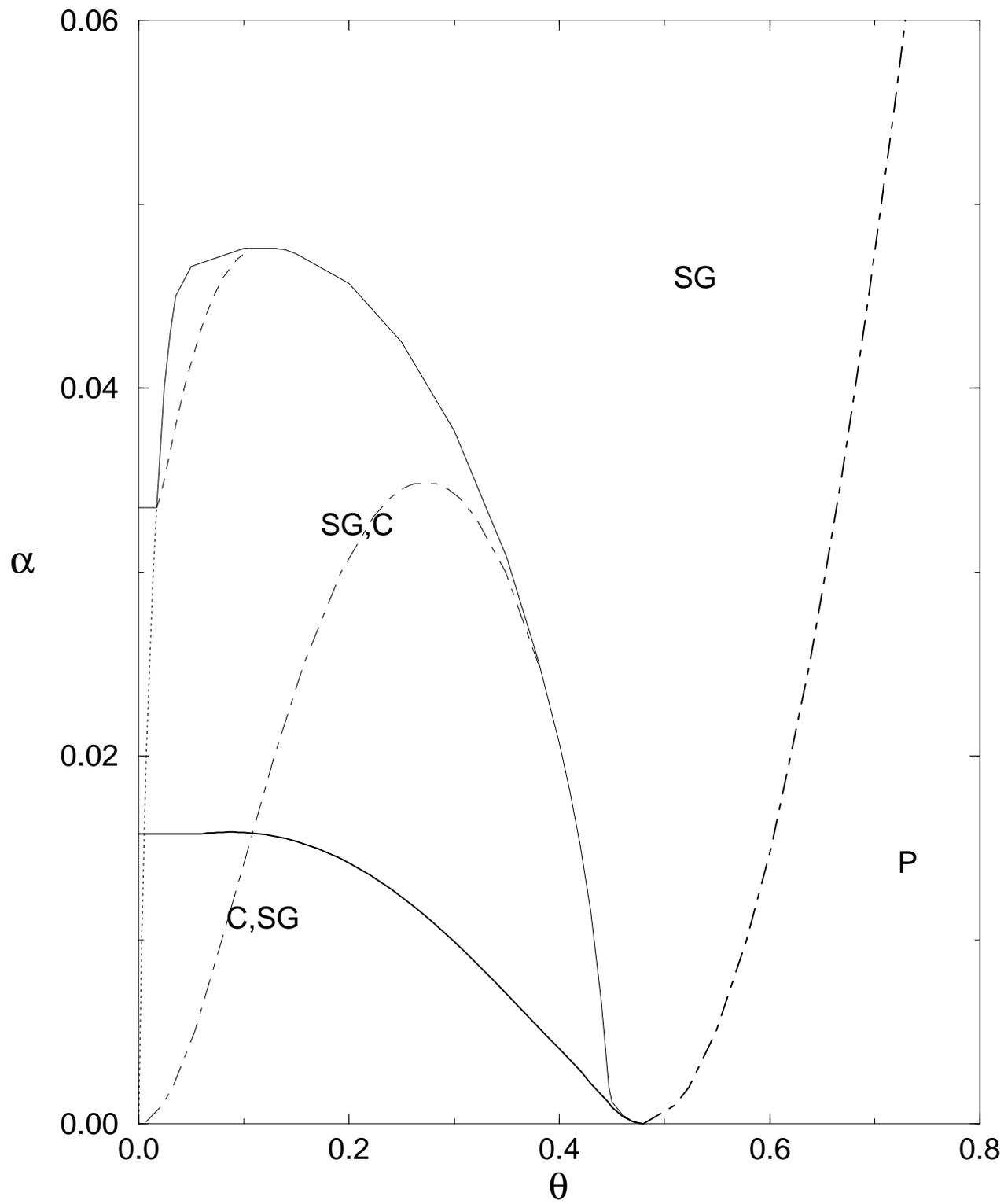
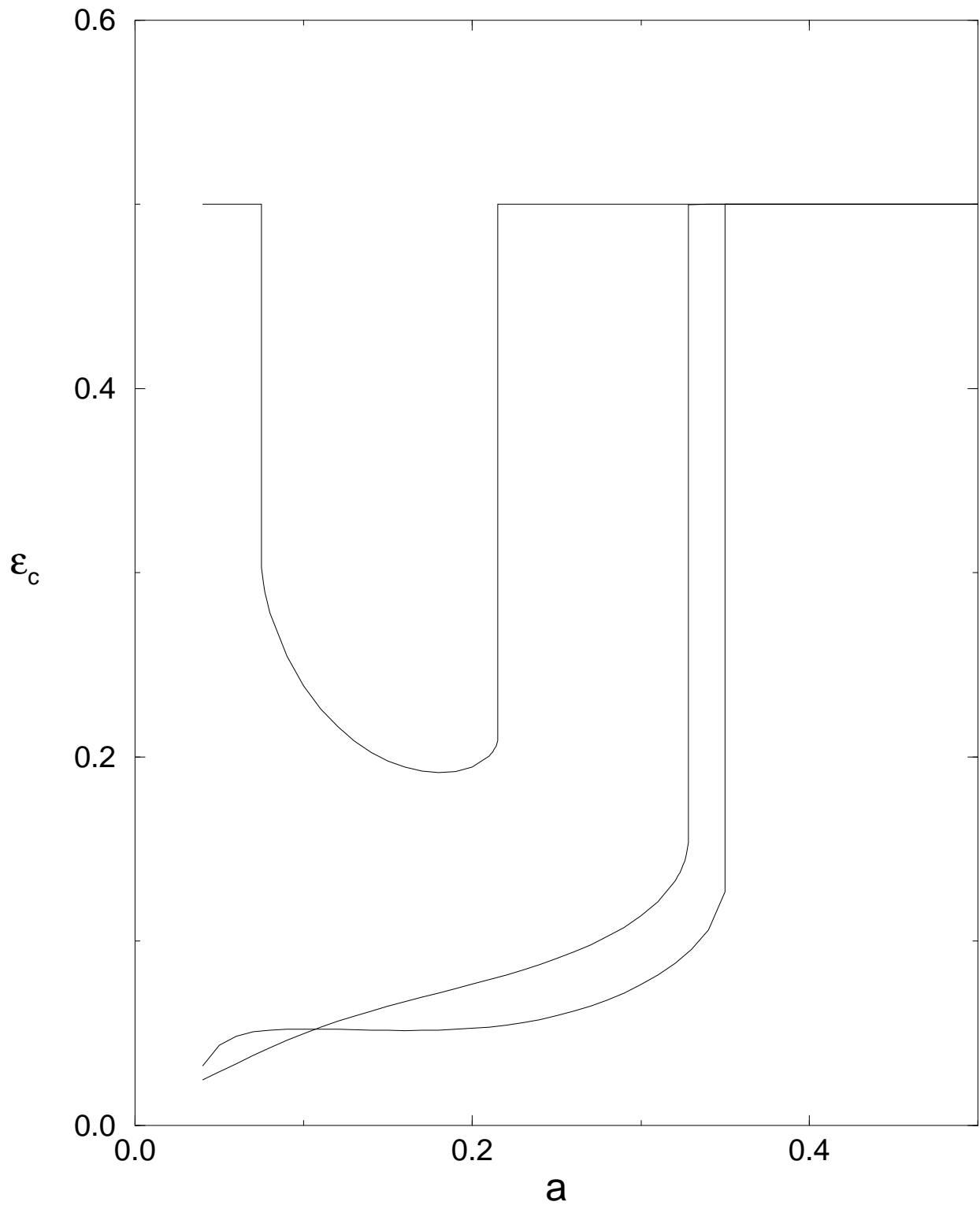
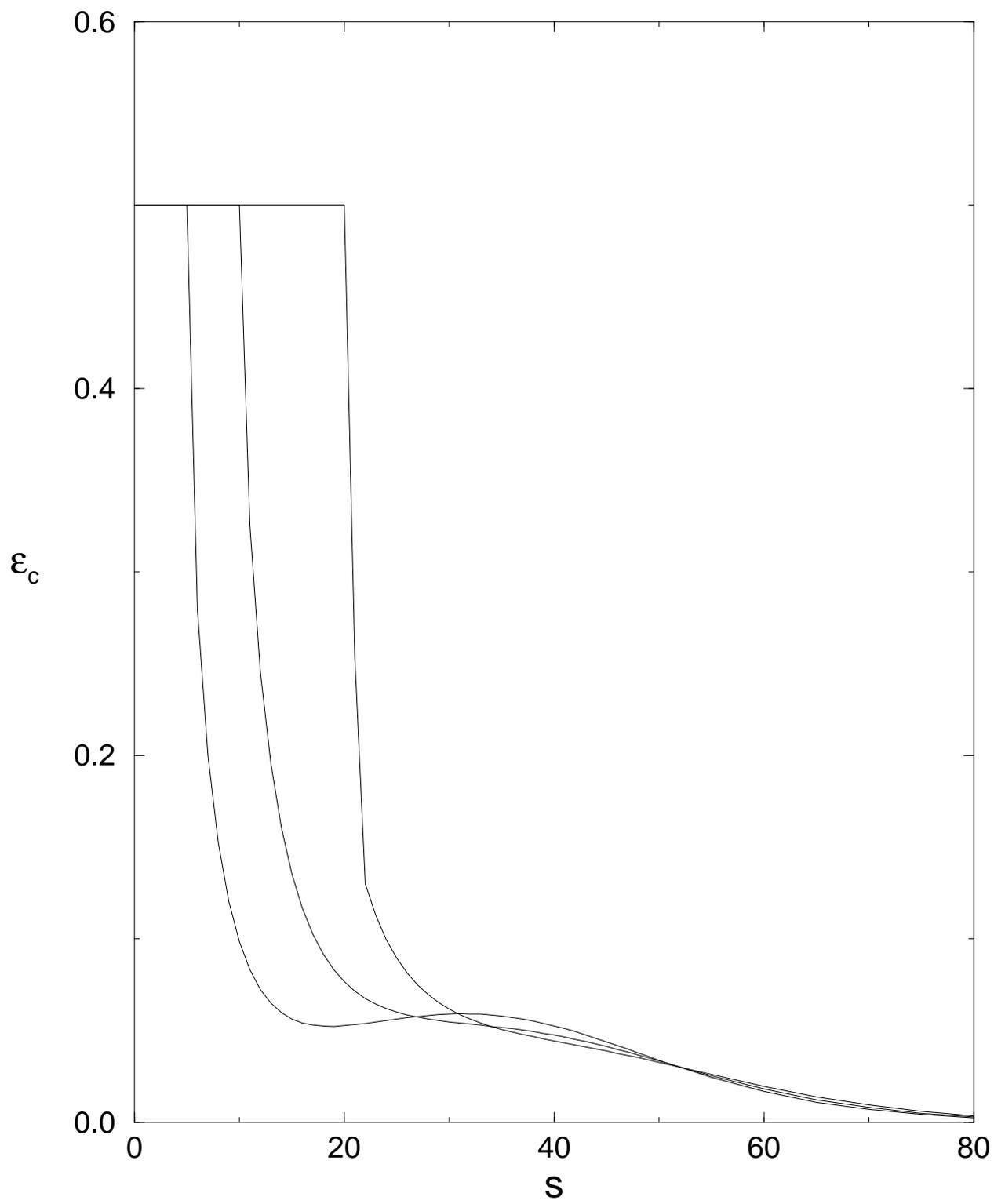
\end{document}